\documentclass{eptcs}

\usepackage{amsmath}
\usepackage{amstext}
\usepackage{amssymb}
\usepackage{amsfonts}
\usepackage{xspace}
\usepackage{graphicx}
\usepackage{url}
\usepackage{subfigure}
\usepackage{mdwlist}
\usepackage{courier}
\usepackage{lscape}
\usepackage{comment}
\usepackage{wrapfig}
\usepackage{multirow}
\usepackage{bussproofs}
\usepackage{pdftricks}
\usepackage{subfigure}
\begin{psinputs}
\usepackage{pstricks}
\usepackage{color}
\usepackage{pstcol}
\usepackage{pst-plot}
\usepackage{pst-tree}
\usepackage{pst-eps}
\usepackage{multido}
\usepackage{pst-node}
\usepackage{pst-eps}
\end{psinputs}
\usepackage{cite}
\makeatletter
\newif\if@restonecol
\makeatother

\usepackage[ruled,vlined]{algorithm2e}
\usepackage{footmisc}
\usepackage{balance}
\usepackage{enumerate}
\usepackage[normalem]{ulem}
\usepackage{hyperref}
\usepackage{alltt}

\newcommand{\mcmt}{{\sc mcmt}\xspace}

\newcommand{\dpll}[1]{{\sc DPLL}\xspace}

\newcommand{\COMMENT}[1]{}

\newcommand{\ui}{\ensuremath{\underline i}}

\renewcommand{\int}{\ensuremath {\mathcal I}}

\newcommand{\tv}{\ensuremath{{\bf v}}\xspace}

\usepackage{boxedminipage}

\title{Monotonic Abstraction Techniques: from Parametric to Software Model Checking}

\author{Francesco Alberti
\institute{University of Lugano, \\Lugano, Switzerland}
\and
Silvio Ghilardi 
\institute{Universit\`a degli Studi di Milano, \\Milano, Italy}
\and
Natasha Sharygina
\institute{University of Lugano, \\Lugano, Switzerland}
}
\def\titlerunning{Monotonic Abstraction}
\def\authorrunning{F. Alberti, S. Ghilardi \& N. Sharygina}

\newcommand{\booster}{{\sc Booster}\xspace}

\begin{document}
%
%
%

%
%
\maketitle              

\begin{abstract}
Monotonic abstraction is a technique introduced in model checking parameterized distributed systems in order to cope with transitions containing global 
conditions within guards. The technique has been re-interpreted in a declarative setting in previous papers of ours and applied to the verification of 
fault tolerant systems under the so-called `stopping failures' model. The declarative reinterpretation consists in logical techniques (quantifier relativizations and,
especially, quantifier instantiations) making sense in a broader context. In fact, we recently showed that such techniques can  over-approximate 
array accelerations, so that they can be employed as a meaningful (and practically effective) component of CEGAR loops in software model checking too.  
\end{abstract}

\section{Introduction}

Monotonic abstraction is a well-known technique introduced by P. A. Abdulla and collaborators in a series of papers (like for instance~\cite{tacas06,cav06,vmcai08,sofsem}); the technique was originally applied 
in the context of verification of distributed systems, but successively  extended also elsewhere (see e.g.~\cite{AbdullaBCHR08,atva09}). 
The approach has been reformulated in~\cite{AlbertiGPRR10a,AlbertiGPRR12} within the declarative context of array-based systems~\cite{GhilardiNRZ08}
in order to apply it to the verification of reliable broadcast algorithms~\cite{toueg} in a fault-tolerant environment. The declarative reformulation makes clear that monotonic abstraction can be viewed operationally as a
purely symbolic manipulation applying quantifier instantiation in order to overapproximate sets of states represented via $\exists^*\forall$-formulae. Since the same kind of formulae arise when computing loop 
accelerations in sequential programs 
for arrays~\cite{AlbertiGS13}, it is natural to import it in software model checking and to merge it with interpolation-based abstraction-refinement techniques~\cite{McMillan06,AlbertiBGRS12a,fmsd}. This strategy has been 
successfully implemented in the latest versions of the model checker \textsc{mcmt} (see~\url{http://users.mat.unimi.it/users/ghilardi/mcmt}). In this contribution, we describe 
the above reinterpretation of monotonic abstraction; we won't enter into technicalities and we shall keep the exposition at intuitive and informal level 
(readers interested in details can consult the published papers  mentioned below). 

\section{Monotonic Abstraction}\label{sec:monotonic}

We quote the content of this Section (a brief description  monotonic abstraction) directly from the tutorial~\cite{rp08}. 

``A parameterized system consists of an arbitrary number number of
processes usually organized as a linear array. In fact, a parameterized system represents an infinite family of systems, namely one for each size of the system. We
are interested in \emph{parameterized verification}, i.e., verifying correctness regardless of
the number of processes inside the system. The term \emph{parameterized} refers to the
fact that the size of the system is (implicitly) a parameter of the verification problem.
Examples of parameterized systems include mutual exclusion algorithms, bus
protocols, telecommunication protocols, and cache coherence protocols.
(...) 

There has been an extensive research on the verification of infinite-state systems which are \emph{monotonic} w.r.t. a \emph{well quasi-ordering} on the set of 
configurations \cite{lics}. The main idea is to perform symbolic backward reachability analysis
to check safety properties for such systems. The method was first reported in~\cite{lossy-lics}
and applied to analyze safety properties for lossy channel systems. Concretely, we
define a pre-order $\preceq$ on the set of configurations such that (1) $\preceq$ is a simulation with
respect to the transition relation (i.e., the transition relation is monotonic w.r.t. $\preceq$),
and (2) $\preceq$
is a well-quasi ordering (WQO for short). Given such a pre-order, we
can derive a backward algorithm for checking reachability of sets of configurations
which are upward closed w.r.t. $\preceq$. Upward closed sets are attractive to use in this
setting for several reasons. First, we are interested in safety properties, in which we
check the reachability of a set of \emph{bad configurations}. These are configurations which
we do not want to occur during the execution of the system. For instance, in mutual
exclusion protocols, the bad configurations are those in which at least two processes
are in their critical sections. This means that checking safety properties amounts to
checking reachability of upward closed sets of configurations. The second attractive
feature of upward closed sets is that they can be characterized by their minimal
elements, which often makes it possible to have efficient symbolic representations of
infinite sets of configurations.

We start from the set of bad configurations, and then compute the sets of predecessors, i.e., sets of configurations which correspond to going one step backwards
along the transition relation. Monotonicity implies that, for any upward-closed set,
the set of its predecessors is an upward-closed set. Since the set of bad configurations
is upward closed, it follows that all the sets which are generated are also upward
closed. This procedure is guaranteed to terminate by the well quasi-ordering of the
relation on the set of configurations.

Since its first application to lossy channel systems~\cite{lossy-lics}, the method has been used
for the design of verification algorithms for a wide range of models such as Petri
nets, timed Petri nets, broadcast protocols, cache coherence protocols, etc. (see,
e.g.,~\cite{icatpn,cav-delzanno,en-lics,bro1}).

Unfortunately, parameterized systems do not quite fit into this framework, in
the sense that there is no nontrivial (useful) WQO for which these systems are
monotonic. The ordering $\preceq$ amounts to the subword relation on words. The main
obstacle is that parameterized systems usually use universal global conditions in
which a process may need to check states of all other processes inside the system.
Universal conditions are inherently non-monotonic, since having larger configurations may lead to the violation of the universal condition. In this tutorial, we give an
overview of the method of \emph{monotonic abstraction}~\cite{tacas06,cav06,vmcai08,AbdullaBCHR08}, which attempts to overcome this problem by defining an abstract semantics which forces monotonicity.
Basically, the idea is to consider that a transition is possible from a configuration $c_1$
to $c_2$ if it is possible from any smaller configuration $c'_1\preceq c_1$ to $c_2$. More precisely, the
abstraction kills (deletes) all the processes inside the configuration which violate the
universal condition. Since the abstract transition relation is an over-approximation
of the original one, proving a safety property in the abstract system implies that
the property also holds in the original system.''

\section{Array-Based Systems}\label{sec:arraybased}

\emph{Array-based systems} were first introduced in~\cite{GhilardiNRZ08}: the underlying idea is that of specifying systems of various nature  using array theories like 
those studied in~\cite{BradleyMS06,BradleyM07} and implemented in common SMT-solvers. Array theories are multi-sorted and very flexible; typically one has three sorts: for indexes, elements and arrays. 
To specify a distributed system, one can use the index sort to model processes and the
 array sort to model the processes locations and more generally local data (like clocks, tickets, etc). Local data and processes locations can in fact be described as functions (aka arrays) mapping indexes to
 suitable elements.
 
 More generally,  an array-based system is specified once a tuple $\tv$ of variables (of index, array or elements sorts) is given,
 together with a formula $I(\tv)$ specifying initialization and with a finite sets of formulae $\tau_h(\tv, \tv')$ specifying system evolution (here the $\tv'$ are renamed copies of the $\tv$, as usual in model-checking
 notation). Below,
 we let $\tau:=\bigvee\tau_h$ and  $\mathsf{Pre}(\tau, K(\tv)):= \exists \tv' (\tau(\tv, \tv') \wedge K(\tv))$; the formula $\mathsf{Pre}(\tau, K(\tv))$ describes the preimage of $K$, i.e. the set of
 states that can reach in one step a state satisfying $K$.
 
 Bad configurations are described via a formula $U(\tv)$. Thus a safety problem becomes a tuple
 \begin{equation}
 \mathcal S=\langle \tv , I(\tv), \{\tau_h(\tv, \tv')\}_{h=1, \dots, t}, U(\tv)\rangle
 \end{equation}
and backward reachability search can be implemented by the algorithm in Figure~\ref{fig:reach-algo}. The algorithm repeatedly computes pre-images of the set of bad states, until either a fixed point is reached 
(the system is `safe') or 
until the intersection with initial states is not empty (the system is `unsafe' in this case, i.e. bad configurations can be reached).
 
 \begin{figure}[tb]
 \begin{center}
 \begin{boxedminipage}{.6\textwidth}
 \begin{tabbing}
 foo \= foo \= \kill
 \textbf{function} $\mathsf{BReach}(\mathcal S)$ \\
 \> $i \longleftarrow 0$; $BR^0(\tau, U) \longleftarrow U$; $K^0 \longleftarrow U$ \\
 \> \textbf{if} $\mathsf{check}(BR^{0}(\tau,U) \wedge I)= \mathsf{sat}$ \textbf{then return} $\mathsf{unsafe}$\\
 \> \textbf{repeat} \\
 \> \> $K^{i+1} \longleftarrow \mathsf{Pre}(\tau, K^i)$\\
 \> \> \textbf{if} $\mathsf{check}(K^{i+1} \wedge I)= \mathsf{sat}$ \textbf{then return} $\mathsf{unsafe}$\\
 \> \> \textbf{else} $i \longleftarrow i+1$\\
 \> \> $BR^{i+1}(\tau, U) \longleftarrow BR^i(\tau, U) \vee K^{i+1}$\\
 \> \textbf{until} $\mathsf{check}(\neg(BR^{i+1}(\tau, U)\to BR^{i}(\tau, U))=\mathsf{unsat}$ \\
 \> \textbf{return} $\mathsf{safe}$ \\
 \textbf{end}
 \end{tabbing}
 \end{boxedminipage}
 \end{center}
 \caption{\label{fig:reach-algo}Backward Reachability for Array-based Systems.}
\end{figure}

Clearly, for the algorithm to be effective, one should be able to discharge the required consistency tests; these tests involve a suitable version $T_{arr}$ of the theory of arrays (the precise identification of this
theory depends on the problem and should be specified together with $\mathcal S$).
We have \emph{safety tests} at lines 3, 6 and \emph{fixpoint tests} at line 9; their effectiveness depends on $T_{arr}$ and on  the shape of
the formulae $I, \tau_h, U$. 

In parameterized distributed systems, there is no arithmetic on index sort (processes are just ordered) and $T_{arr}$ is quite simple (see~\cite{GhilardiR10a}); in this situation, 
if we view the system variables $\tv$ as fresh constants,
configurations can be 
identified with finitely generated models of $T_{arr}$ (with generators having all index sort), ordering among configurations is model-theoretic embeddability and upward sets are characterizable via definability with existential sentences
(i.e. sentences obtained by prefixing a string of existential index quantifiers to quantifiers-free formulae). We thus obtain a reformulation of the framework of well-structured systems~\cite{lics} in a purely
logical context, using basic model-theoretic ingredients.

Now, if $U$ is existential and if the preimage of an existential sentence is existential, we always generate existential sentences in backward search. Since usually the initial 
formula $I$ is universal (i.e. its negation is existential),
fixpoint and safety tests only require testing consistency of the conjunction of an existential sentence with a universal one; this is  decidable if $T_{arr}$ is sufficiently simple. 
So \emph{the only problem is to guarantee that the preimage of an existential formula along a transition $\tau_h$ is still existential}. Here we re-encounter (mutatis mutandis) the problem of universal conditions in guards.

The problem does not arise if the transition formulae are guarded assignments in purely functional forms 
\begin{equation}\label{eq:functional}
 \exists \ui~(\phi(\ui, \tv) \wedge \tv'= F(\ui, \tv))
\end{equation}
Here $\ui$ is a tuple of existentially quantified index variables, $\phi$ is quantifier-free and $F$ is a quantifier-free definable function (we do not enter into technical details, but we point out that with such $F$
you can express array updates making use of case-distinctions). Now, transitions like~\eqref{eq:functional} can easily express e.g. broadcast and synchronizations actions,  but they are insufficient to formalize for instance a 
basic `bakery' mutual exclusion protocol. To formalize the latter, one needs universal quantifiers in guards, i.e. we need the following  format
\begin{equation}\label{eq:functional1}
 \exists \ui~(\phi(\ui, \tv) \wedge \forall k \,\psi(k, \ui, \tv) \wedge \tv'= F(\ui, \tv))
\end{equation}
as shown in the example below.

{\small
\vskip 2mm
\noindent \textbf{Example.}
We consider a protocol (taken
from~\cite{sofsem}) ensuring mutual exclusion
for an arbitrary number of processes with a linear topology.  Each
process has four control locations: I(dle), R(equesting), W(aiting),
and C(ritical section).
The processes are linearly ordered: the set
of processes on the left (right) of a given process $p$ are those
processes coming before (after, respectively)  $p$ in the linear
order.  Initially, all  processes are in control location I.  Each
process can perform the following transitions:
\begin{description}
\item[$t_1$:] if a process is in location I and all other processes
  (on its left and right) are either in location I or R, then it can
  move to location R;
\item[$t_2$:] if a process is in location R, then it can move to
  location W;
\item[$t_3$:] if a process is in location W and all the processes on
  its left are in location I, then it can move to location C;
\item[$t_4$:] if a process is in location C, then it can move to
  location R; and
\item[$t_5$:] if a process is in location R, then it can move to
  location I.
\end{description}
All those (bad) states containing at least two distinct processes in
the control location C violate the mutual exclusion property that the
protocol is intended to guarantee.

To formalize this protocol as an array-based system, we use as $\tv$ the single array variable $a$;
the theory $T_{arr}$ is a very weak fragment of standard array theories; we
specify $I, \tau_h, U$ below.

The set of initial states is characterized by
\begin{eqnarray*}
  I(a) := \forall z.a[z]=\mathsf{I}, 
\end{eqnarray*}
and the transitions are formalized by
\begin{eqnarray*}
  \begin{array}{lcll}
  \tau_1(a,a') & := & \exists i.(a[i]=\mathsf{I}~~~~~ \wedge ~~~
                      \forall j.(j\neq i\Rightarrow 
                                 a[j]\in \{\mathsf{I}, \mathsf{R}\}) ~~~\wedge
                 & a'=\mathtt{upd}(a,i,\mathsf{R})) \\
  \tau_2(a,a') & := & \exists i.(a[i]=\mathsf{R} ~~~~\wedge
                 & a'=\mathtt{upd}(a,i,\mathsf{W})) \\
  \tau_3(a,a') & := & \exists i.(a[i]=\mathsf{W}~~\wedge ~~~
                      \forall j.(j< i\Rightarrow a[j]=I) \wedge
                 & a'=\mathtt{upd}(a,i,\mathsf{C})) \\
  \tau_4(a,a') & := & \exists i.(a[i]=\mathsf{C} ~~~~\wedge
                 & a'=\mathtt{upd}(a,i,\mathsf{R})) \\
  \tau_5(a,a') & := & \exists i.(a[i]=\mathsf{R}~~~~ \wedge
                 & a'=\mathtt{upd}(a,i,\mathsf{I})) ,
  \end{array}
\end{eqnarray*}
where 
$a[j]\in \{\mathsf{I}, \mathsf{R}\})$ abbreviates 
$a[j] =\mathsf{I} \vee a[j]= \mathsf{R}$
 and
$\mathtt{upd}(a,i,c)$ abbreviates 
the case-defined function 
$\lambda j.(\mathtt{if}~(j=i)~\mathtt{then}~c~\mathtt{else} 
~a[j])$ 
for
$a,i$, and $c$ constants of appropriate sorts. 
Notice that $\tau_2, \tau_4, \tau_5$ have the form~\eqref{eq:functional} but $\tau_1, \tau_3$ have the form~\eqref{eq:functional1}.

  The formula describing the set of bad states is 
\begin{eqnarray*}
  U(a) & := & \exists i,j.(i < j\wedge a[i]=\mathsf{C}\wedge 
                           a[j]=\mathsf{C}) .
\end{eqnarray*}
Notice that $U$ is existential. $\hfill \dashv$
\vskip 2mm
}

Since the preimage along transitions having the form~\eqref{eq:functional1} do not yield existential formulae, we modify them by monotonic abstraction. 
The idea is to re-interpret monotonic abstraction in the following way (which seems to be loosely motivated by the `stopping failures' paradigm in distributed algorithm literature~\cite{lynch}).
We assume that processes can crash at any instant of time and that crashed processes do not take part anymore to the protocol (in the terminology of Section~\ref{sec:monotonic} we are removing them, by
passing to a `sub-configuration'). In this setting,  a transition of  the form~\eqref{eq:functional1} can always fire, provided the processes violating the universal guard $\forall k \,\psi(k, \ui, \tv)$ crash.
This transformation can be interpreted as a modification of the underlying computational model (we are adopting the `stopping failures' paradigm) or 
more simply just as a kind of abstraction. One should be aware that the modified 
system has more runs, so safety of the modified system implies safety of the original one but not vice versa. Whatever it is, the point is that in the context of array-based system, 
\emph{this monotonic abstraction modification can be performed
at the syntactic level}: by using quantifier relativizations and  by adding a `crash' case to the update function $F$, it is possible to transform a transition $\tau_h$ having 
the form~\eqref{eq:functional1} into a transition ${\hat\tau_h}$ having the form~\eqref{eq:functional} (details are fully explained in~\cite{AlbertiGPRR12}).

From the experimental viewpoint, the above approach was successfully implemented in the tools \textsc{mcmt} and \textsc{safari}~\cite{ABGRS12b}.
Since its first releases, \textsc{mcmt} showed very good results in the verification of several classes of benchmarks taken from heterogeneous application domains (timed automata, broadcast protocols, cache coherence protocols).
Two significant case studies \cite{BruttomessoCGR12, AlbertiGPRR10a} provided empirical evidence of the effectiveness of this approach on non-trivial problems.
We acknowledge also a more recent re-implementation of \textsc{mcmt} on a parallel architecture in the tool \textsc{cubicle}~\cite{cubicle1}. With the help of further advanced heuristics, 
\textsc{cubicle} outperformed previous tools and obtained impressive results~\cite{cubicle2} (including the first completely automatic parameterized verification of the `Flash' cache coherence protocol).
Broadening the horizon of possible applications, the \textsc{mcmt} approach has been adopted in \cite{AlbertiAR11a, AlbertiAR11b} to check the absence of flaws in Role-Based Access Control policies.

\section{Array Acceleration}

Let us examine more closely the effect of the syntactic monotonic abstraction of the previous section. If we take an existential formula
$K$ and a transition $\tau_h$ of the form~\eqref{eq:functional1}, the preimage $ \mathsf{Pre}(\tau_h, K)$ has the form
\begin{equation}\label{eq:pre}
 \exists \ui\, \forall k\, \psi(\tv, \ui, k),
\end{equation}
where $\psi$ is quantifier-free.
This formula can be overapproximated by an existential formula by taking
\begin{equation}\label{eq:pre1}
 \exists \ui\, \bigwedge_t \psi(\tv, \ui, t),
\end{equation}
varying $t$ among a set of terms $X$. We may call~\eqref{eq:pre1} a \emph{monotonic abstraction of the formula}~\eqref{eq:pre} (notice that this notion is relative to $X$).
There is an interesting relationship between the syntactic monotonic abstraction for transitions introduced in the previous section and the above monotonic abstraction for formulae.
Indeed, if one take the obvious choice $X:=\ui$, it turns out that $ \mathsf{Pre}(\hat\tau_h, K)$ is precisely~\eqref{eq:pre1}. Thus, monotonic abstraction can be applied at run-time (i.e. during backward search)
and not in a preprocessing step  (replacing the $\tau_h$ by the corresponding $\hat \tau_h$). This observation makes the technique of monotonic abstraction very flexible:
one may choose $X$ depending on the formula~\eqref{eq:pre}, according to
some heuristics.

Let us now explain how monotonic abstraction can be imported in software model checking. First, notice that we can easily translate a sequential program manipulating integers and array variables
into an array-based system. We get some simplifications with respect to distributed systems: first, initial and unsafe formulae are quite simple, they just say that a specific integer variable
$pc\in \tv$ (the `program counter') is equal to the initial and to the error locations, respectively. There are no quantifiers involved here and there are no quantifiers in transitions too. In fact, transitions are
ground assignments of the form 
\begin{equation}\label{eq:functional0}
 \phi(\tv) \wedge \tv'= F(\tv)
\end{equation}
corresponding to the various instructions of an imperative program.
Fixpoint and safety tests consequently involve just ground formulae, so they are easily seen to be decidable. 

However, the  array theory $T_{arr}$ needed is now more complicated, because we have some arithmetic on indexes
(usually we need Presburger arithmetic, but often difference logic is sufficient). The complications in the array theory compromise termination analysis, because it is not possible anymore to get a WQO among configurations
(whatever `configuration' may mean here). In fact, termination is the major source of problems, even from the practical point of view. Let us show what happens in an example.

{\small
\vskip 2mm
\noindent \textbf{Example.}
The following `initialize-and-test' simple example is quoted as problematic for CEGAR techniques in~\cite{JhalaM07}:
\vskip 1mm
\centerline{
{\tt for(I=0; I!= a\_length; I++) a[I]=0;~~~~~~~~~}
}

\centerline{
{\tt for(J=0; J!= a\_length; J++) assert(a[J]==0);}
}
\vskip 1mm
\noindent
We first translate the above pseudo-code into  the formalism
of array-based systems. We need two integer  variables $I,J$, a program counter $p$ and an array variable
$a$. We have five transitions:

$$
\begin{aligned}
& \tau_0 = \left(
\begin{aligned}
&~~~~~ p = 0~ \wedge~
 p' = 1 ~\wedge \\
&  I' = 0~ \wedge~ J'=J ~\wedge ~a'=a;
\end{aligned}
\right) \\
& \tau_1 = \left(
\begin{aligned}
& p = 1 \wedge I \neq a\_length  ~\wedge~
 p' = 1 ~\wedge \\
&  I' = I+ 1~ \wedge~ J'=J ~\wedge ~a'=wr(a,I,0);
\end{aligned}
\right) \\
& \tau_2 = \left(
\begin{aligned}
& p = 1 \wedge I = a\_length  ~\wedge~
 p' = 2 ~\wedge \\
&  I' = I ~\wedge~ J'=0~ \wedge~ a'=a;
\end{aligned}
\right) \\
& \tau_3 = \left(
\begin{aligned}
p = 2 &~\wedge~ J \neq a\_length  ~\wedge~ a[J]=0 \\
 p' = 2 ~\wedge ~
&  I' = I ~\wedge~ J'=J+1~ \wedge~ a'=a;
\end{aligned}
\right) \\
& \tau_4 = \left(
\begin{aligned}
p = 2 &~\wedge~ J \neq a\_length  ~\wedge~ a[J]\neq 0 \\
 p' = 4 ~\wedge ~
&  I' = I ~\wedge~ J'=J~ \wedge~ a'=a;
\end{aligned}
\right) \\
& \tau_5 = \left(
\begin{aligned}
&p = 2 ~\wedge~ J = a\_length  ~\wedge ~
 p' = 3 ~\wedge \\ 
&  I' = I ~\wedge~ J'=J~ \wedge~ a'=a;
\end{aligned}
\right) 
\end{aligned}
$$
where $wr(a,i,e)$ represents a copy of the array $a$ with the element $e$ written in position $i$.

The system is initialized by $p=0$ and the unsafe condition is unreachability of location 4, namely it is represented by the formula $p=4$.

Backward search trivially diverges here, because it produces formulae like
$$
\begin{aligned}
& p=2 ~\wedge ~ J \neq a\_length ~\wedge a[J]\neq 0
\\
& p=2 ~\wedge ~ J+1 \neq a\_length ~ \wedge~ a[J+1]\neq 0~\wedge a[J]= 0
\\
& ~~~\cdots \\
& p=2 ~\wedge ~
J+n \neq a\_length ~\wedge a[J+n]\neq 0~ \wedge~\bigwedge_{k=J}^{J+n-1} a[k]=0
\\
& ~~~\cdots \\
\end{aligned}
$$
$\hfill\dashv$
\vskip 2mm
}

To stop divergence, we need to re-introduce  quantifiers. One possible solution is to summarize the effect of $n$ executions of a loop into a single transition, 
representing transitive closure.
This technique is known as \emph{acceleration} in model-checking and has been extensively investigated for fragments of Presburger arithmetic: integer relations
having definable transitive closure 
include 
relations that can be formalized as
difference bounds constraints~\cite{db1,db2}, octagons~\cite{octagons}
and finite monoid affine transformations~\cite{presburger}
(the paper~\cite{acceleration_CAV} presents a general approach covering all
these domains).

In our setting, we need acceleration inside array theories: the subject is investigated in~\cite{AlbertiGS13}, where a class of ``acceleratable'' ground guarded assignments is identified.
The crucial observation is that \emph{accelerated transitions have the form~\eqref{eq:functional1}}, hence syntactic monotonic abstraction can be applied here.
In the example above, we can accelerate transitions 1 and 3, resulting in
$$
\begin{aligned}
& \tau_1^+ = \exists n>0 ~\left(
\begin{aligned}
& p = 1 \wedge \forall k~(I\leq k< I+n \to k \neq a\_length)  ~\wedge~
 p' = 1 ~\wedge \\
&  I' = I+ n~ \wedge~ J'=J ~\wedge ~a'=wr(a,[I, I+n-1],0)
\end{aligned}
\right); \\
& \tau_3^+ = \exists n>0~\left(
\begin{aligned}
&p = 2 ~\wedge~ \forall k~(J\leq k<J+n \to k \neq a\_length  ~\wedge~ a[k]=0) \\
&\wedge~ p' = 2 ~\wedge ~
  I' = I ~\wedge~ J'=J+n~ \wedge~ a'=a
\end{aligned}
\right).
\end{aligned}
$$
Applying acceleration and syntactic monotonic abstraction, non-termination can often be avoided: for instance, the `initialize-and-test' problem above is solved easily
and correct quantified invariants are synthesized.

Some remarks are in order, however, to point out some differences with respect to the framework of Section~\ref{sec:arraybased}.
First of all, the array theory $T_{arr}$ used in  this section is more powerful and as a consequence satisfiability tests are effective
only
for more restricted classes of formulae.
A decidable class include existential formulae and this class is sufficient to discharge safety tests (because the initial formula  is ground now - it just says that the program counter is
initialized to the first line code location). For fixpoint tests, the situation is different because here we must test for satisfiability the conjunction of an existential formula with a universal one. 
The idea is to use some instantiation-based incomplete calculus; the effect is not tremendous,
i.e., it does not compromise the soundness of the tool,
because if the model-checker fails to recognize a fixpoint, it is subject to extra work and possibly to
divergence, but it does not produce wrong safe/unsafe outputs for this reason.

There is also a good new, however. Monotonic abstraction is just a heuristics for abstraction here, it may not produce spurious answers, if handled carefully. In fact, the addition of accelerated transitions does not
modify the semantics of the system (contrary to what happened with the adoption of the stopping failure model), 
so error traces containing accelerated transitions can simply be ignored (if the system is unsafe, there should be a way to discover it without using accelerated transitions). 
So the best way to exploit monotonic abstraction for accelerated transitions is to use this technique inside abstract/refinement cycles, as an additional machinery to produce abstractions.

\begin{table}[t]
\begin{tabular}{||l|r|r|r|r|r|r|r|r|r||}
\hline\hline
Problem &kind & d & \#n & \#del & \#SMT  &\#inv &\#ref  &heur &time 
             \\
\hline\hline
Illinois  &(C) 	 &4  &8  &0  &212  &0 	  &0    & - &0.06                   \\ \hline
German  &(C) 	 &26  &2121  &255  &117121  &0	  &0    & - &60.76                   \\ \hline
German\_buggy  &(C) 	 &16  &1300  &203  &24275  &0 	  &0    & - &14.28                   \\ \hline
Bakery  &(M) 	 &2  &1  &0  &29  &0 	  &0    & - &0.00                   \\ \hline
Szymanski  &(M) 	 &11  &17  &5  &1092  &12 	  &0    & I &0.21                   \\ \hline
Szymanski\_atomic  &(M) 	 &19  &63  &7  &5470  &32 	  &0    & I &1.82                   \\ \hline
Distributed Lamport  &(M) 	 &23  &248  &42  &19622  &7 	  &0    & I &27.18                   \\ \hline
Crash &(D) 	 &13  &113  &21  &1731  &0 	  &0    & - &0.75                   \\ \hline
Fischer &(D) 	 &10  &16  &2  &363  &0 	  &0    & - &0.08                   \\ \hline
Fischer\_buggy &(D) 	 &6  &16  &0  &307  &0 	  &0    & - &0.06                   \\ \hline
Lynch-Shavit\_full &(D) 	 &25  &1103  &99  &56638  &0 	  &0    & - &33.39                   \\ \hline
Strcpy &(S) 	 &4  &4  &2  &48  &0 	  &0    & A &0.01                   \\ \hline
Strcmp &(S) 	 &6  &10  &4  &128  &0 	  &0    & A &0.02                   \\ \hline
Max\_in\_array &(S) 	 &7  &13  &6  &166  &0 	  &0    & A &0.04                   \\ \hline
Reverse &(S) 	 &4  &8  &5  &101  &0 	  &0    & A &0.03                   \\ \hline
Palindrome &(S) 	 &4  &7  &4  &107  &0 	  &0    & A &0.04                   \\ \hline
AllDifferent &(S+) 	 &7  &49  &39  &871  &0 	  &8    & A\! +\! AR &0.40                   \\ \hline
BubbleSort &(S+) 	 &5  &14  &10  &200  &0 	  &0    & A &0.07                   \\ \hline
InsertionSort &(S+) 	 &18  &98  &56  &3874  &0 	  &2    & AR &1.43                   \\ \hline
SelectionSort &(S+) 	 &8  &101  &77  &6059  &8 	  &11    & AR\! +\! I &4.98                   \\ \hline
\end{tabular}
\\
\caption{\textsc{mcmt} Statistics.~{\small The experiments were run on a laptop
Intel(R) Core(TM) i3 CPU   \@ 2.27GHz with 4GB RAM 
 running Linux  Ubuntu 12.04.
In the second  column we indicate the class of the problem: (M) mutual exclusion, (C) cache coherence, (D) other distributed
protocols (timed, fault tolerant, etc.), (S) sequential program for arrays, (S+) sequential program for arrays with nested loops.
In  column 3-8 we respectively give the depth of the search tree, the number of nodes generated by the tool in the search tree, 
the number of subsumed or subcovered nodes, the number of calls to the SMT solver, the number of invariants found by the tool in forward search 
and the number of refinements
applied in abstraction/refinement mode. In the last column we put
the total time in seconds and in the last-but-one column a summary of the options used (A=acceleration, AR=abstraction/refinement, 
I=invariant search). 
For each problem we reported the result in the best configuration we found for the tool.}}\label{tab:mcmt}
\end{table}

\section{Implementations}

A framework for abstraction/refinement based on interpolation was introduced in~\cite{AlbertiBGRS12a,fmsd}
as an extension to array programs of McMillan technique~\cite{McMillan06}; the framework was first implemented in the \textsc{safari} tool~\cite{ABGRS12b}. The model-checker \textsc{mcmt}, since version 2.0, implements 
both abstraction/refinement for arrays and array acceleration with monotonic abstraction. Merging the two techniques has considerably raised the number of benchmarks solved by the tool. 

The \textsc{mcmt} architecture is rather complex, because it contains not only the interface with the underlying SMT-solver (which is \textsc{Yices}), but also various modules for acceleration, abstraction, quantifier
instantiation and quantifier elimination. We refer the reader to~\cite{cilc14}, for a presentation of the recent version 2.5. Table~\ref{tab:mcmt} below summarizes some performances on well-known benchmarks.
We underline that the verification problems are all solved in their \emph{parametric} version, i.e. regardless to the dimension of the system/length of arrays or strings.

It should be noticed  that
often array acceleration produces quantified formulae that falls within decidable classes like those studied in~\cite{BradleyMS06,AlbertiGS14}; 
thus, if the control flow of the program is flat, it is not convenient to apply monotonic abstraction. 
This is what happens inside \booster\footnote{Available at \url{www.inf.usi.ch/phd/alberti/prj/booster}.}.
%
\booster is an integrated framework providing, among other static
analysis techniques, an implementation of the acceleration paradigm
presented in \cite{AlbertiGS14}, where acceleration is exploited
\emph{precisely} without the adoption of the monotonic abstraction
technique. \booster implements also heuristics to execute \mcmt in a
parallel way, in order to try different promising configurations of the
tool.

\paragraph{\bf Acknowledgements.} The work of the first author was
supported by the Swiss National Science Foundation under grant no.
P1TIP2 152261 and the one of the second by Italian Ministry of
Education, University and Research (MIUR) under the PRIN 2010-2011
project ``Logical Methods for Information Management''.

\bibliographystyle{eptcs}
\bibliography{references}

\end{document}